# Successive phase transitions of the spin–orbit-coupled metal $Cd_2Re_2O_7$ probed by high-resolution synchrotron X-ray diffraction

Daigorou Hirai,[1,2 a)] Atsuhito Fukui,[1] Hajime Sagayama,[3] Takumi Hasegawa,[4] and Zenji Hiroi[1]

[1]Institute for Solid State Physics, University of Tokyo, Kashiwa, Chiba 277-8581, Japan
[2]Department of Applied Physics, Nagoya University, Nagoya 464–8603, Japan
[3]Institute of Materials Structure Science, High Energy Accelerator Research Organization, Tsukuba, Ibaraki 305-0801, Japan
[4]Graduate School of Advanced Science and Engineering, Hiroshima University, Higashi-Hiroshima 739-8521, Japan

E-mail: dhirai@nuap.nagoya-u.ac.jp

**Abstract**

The $5d$ pyrochlore oxide superconductor $Cd_2Re_2O_7$ (CRO) has attracted significant interest as a spin–orbit-coupled metal (SOCM) that spontaneously undergoes a phase transition to an odd-parity multipole phase by breaking the spatial inversion symmetry due to the Fermi liquid instability caused by strong spin–orbit coupling. Despite the significance of structural information during the transition, previous experimental results regarding lattice deformation have been elusive. We have conducted ultra-high resolution synchrotron radiation X-ray diffraction experiments on a high-quality CRO single crystal. The temperature-dependent splitting of the 0 0 16 and 0 0 14 reflections, which are allowed and forbidden, respectively, in the high-temperature cubic phase I (space group $Fd$–$3m$), has been clearly observed and reveals the following significant facts: inversion symmetry breaking and tetragonal distortion occur simultaneously at $T_{s1} = 201.5(1)$ K; the previously believed first-order transition between phase II ($I$–$4m2$) and phase III ($I4_122$) at $T_{s2} \sim 120$ K consists of two close second-order transitions at $T_{s2} = 115.4(1)$ K and $T_{s3} \sim 100$ K; there is a new orthorhombic phase XI ($F222$) in between. The order parameters of these continuous transitions are uniquely represented by a two-dimensional irreducible representation $E_u$ of the $O_h$ point group, and the order parameters of phase XI are a linear combination of those of phases II and III. Each phase is believed to correspond to a distinct odd-parity multipole order, and the complex successive transitions observed may be the result of an electronic phase transition that resolves the Fermi liquid instability in the SOCM.

## 1. Introduction

In general, the Fermi surfaces of metals become unstable against various interactions, and characteristic ordered states appear to resolve them. When electron–phonon and electron–electron interactions become strong, for instance, a gap opens on the Fermi surface, and ordered states such as superconductors and Mott insulators, respectively, are produced. Behind observed phase transitions are interactions that govern the physical properties, and the investigation of ordered states leads to an in-depth comprehension of these interactions.

Recently, Fu proposed the concept of spin–orbit-coupled metal (SOCM), taking into account the rarely considered Fermi liquid instability resulting from spin–orbit couplings (SOCs).[1] The SOCM possesses a crystal structure with inversion symmetry and strong SOCs that act on the conduction electrons. There, the SOC-caused Fermi liquid instability induces spontaneous inversion symmetry breaking (ISB). As a result, the antisymmetric SOC is activated, and spin splitting is expected to occur on the Fermi surface, resulting in odd-parity ordering of itinerant electrons such as multipoles, gyrotropic order, and ferroelectric metallic phases.[1] Conversely, ISB occurs to resolve the instability caused by SOC on spin-degenerate Fermi surfaces. Furthermore, it has been proposed that fluctuations in odd-parity multipole orders could induce exotic *p*-wave superconductivity.[2]

$Cd_2Re_2O_7$ (CRO) has garnered considerable interest as a promising SOCM candidate.[3] It satisfies the requirements for SOCM due to its cubic pyrochlore-type crystal structure with inversion symmetry at room temperature and its metallic conductivity with $5d$ electrons that have strong SOCs.[4] In fact, it exhibits spontaneous ISB. As shown in Fig. 1, phase I at room temperature has a cubic structure with the space group $Fd$–$3m$, while phase II below $T_{s1} \sim 200$ K has a tetragonal structure with the space group $I$–$4m2$ and broken inversion symmetry.[5] $T_{s1}$ is a second order transition with negligible tetragonal deformation between 0.05-0.10 percent.[6,7] Despite this minor structural modification, the electronic state changes drastically below $T_{s1}$, with a sharp decrease in electrical resistivity and a nearly 50 percent decrease in density of states.[3] Therefore, $T_{s1}$ is considered an electronic phase transition driven by the Fermi liquid instability of SOCM, as opposed to a simple structural transition.[1,3]

In addition, it is believed that phase II transitions to phase III with a different tetragonal structure in the space group $I4_122$ and broken inversion symmetry below $T_{s2} \sim 120$ K; this transition is reported to be a first-order transition.[8] Under high pressure, on the other hand, a total of six phases, up to phase IX, have been identified in the vicinity of the critical pressure of 4.2 GPa for ISB,[9] and an $R$–$3m$ phase appears at room temperature above 21 GPa;[10] we refer to this phase as phase X. These diverse phases indicate that CRO possesses a unique Fermi liquid instability and a coupled structural instability.

The two low-temperature phases of ISB at ambient pressure most likely correspond to the multipole order with odd parity predicted for SOCM.[1] The Landau theory explains the successive phase transitions between phases I ($Fd$–$3m$), II ($I$–$4m2$), and III ($I4_122$) in terms of the order parameters (OPs) of the $E_u$ irreducible representation of the $O_h$ point group.[12] As depicted in Fig. 1, the displacements induced by phase transitions in the four tetrahedral Re atoms can be viewed as electric dipoles, certain pairs of which generate electric toroidal moments. Then, for phases II and III, these virtual electric toroidal moments are organized as $x^2 - y^2$ and $3z^2 - r^2$ configurations, respectively.[11] They are therefore known as electric toroidal quadrupole (ETQ) orders.[13] Alternatively, they can be interpreted as the distinct orderings of the six Re–Re bonds in the tetrahedron.[13]

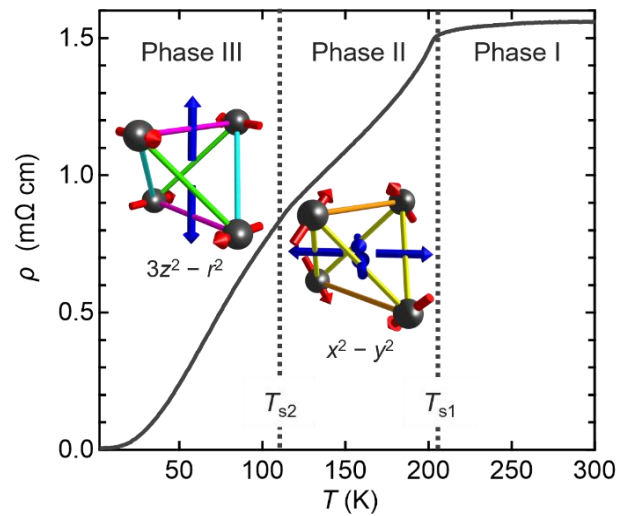

Figure. 1. Temperature dependence of electrical resistivity for a high-quality $Cd_2Re_2O_7$ crystal with RRR = 670. There is a distinct kink at $T_{s1}$, where space inversion symmetry is broken in a second order manner, whereas a broad anomaly appears around $T_{s2}$. The inserted images depict how the Re tetrahedron can deform in phases II and III, with red arrows representing the displacements of Re atoms (black balls).[11] The Re displacements generate virtual electric toroidal moments (blue arrows) of the $x^2 - y^2$ and $3z^2 - r^2$ types, respectively. The colors of the connecting rods between Re atoms differentiate identical bonds in each phase.

Experimental results contradicting the aforementioned were also reported for the successive structural phase transitions of CRO. Single crystal X-ray diffraction (XRD),[5,14] powder neutron diffraction,[7] convergent-beam electron diffraction (CBED),[15] and Raman scattering experiments[16] support the space group $I$–$4m2$ for phase II, whereas nonlinear optical second harmonics generation (SHG) experiments suggest a less symmetric $I$–$4$;[17] an alternative OP to $E_u$ has been discussed.[17,18] For phase III, the $I4_122$ space group is supported by single crystal XRD[5,14] and CBED,[15] whereas the $T_{s2}$ transition was not observed in SHG measurements.[19]

Recent Raman scattering experiments with an isotope-substituted CRO crystal demonstrated that a third phase transition occurs at an even lower temperature of 80 K.[20] According to first-principles calculations, an orthorhombic phase of the space group $F222$ is stable down to the lowest temperature.[20] However, the veracity of this assertion is questionable, given that no structural or physical anomalies at 80 K have been observed in other measurements, including other Raman scattering experiments.[16,21] Very recently, on the other hand, Takigawa observed in his Cd NMR experiments that $T_{s2}$ is not a first-order transition and that a new phase with low symmetry exists between phases II and III.[22] Furthermore, Uji et al. obtained compatible torque measurement results.[23]

We performed high-resolution synchrotron radiation XRD experiments on a high-quality CRO crystal to elucidate the details of structural changes caused by the phase transitions; recent improvements in crystal quality resulted in a one-order-of-magnitude increase in residual resistivity ratio (RRR), indicating less carrier scattering by defects, and allowed the observations of quantum oscillations as well as the flipping of tetragonal strain at $T_{s2}$.[11,24,25] Consequently, we were able to observe the distinct splitting of diffraction peaks caused by tetragonal distortion, which had not been observed in previous XRD experiments.[6] We determined the precise temperature dependence of the lattice constants and gathered data on the extinction that uniquely identifies the space group for each phase. The most significant discovery is that the 120 K transition is not a first-order transition but rather consists of two close second-order transitions at $T_{s2}$ = 115.4(1) K and $T_{s3}$ ~ 100 K, with the formation of a new orthorhombic phase XI overlooked for years in between. Unique structural modifications unveiled in this study, such as the negative thermal expansion below $T_{s1}$ and the enhancement of crystal symmetry at $T_{s3}$, strongly suggest that the electronic phase transition is the source of these structural modifications. As a result of these observations, it is anticipated that our understanding of the microscopic origin of the odd-parity multipoles in CRO and the characteristics of SOCM will advance.

## 2. Experiment

Synchrotron XRD experiments were performed at the beamline BL-4C of the KEK Photon Factory. We measured the temperature dependence of two high-index (0 0 16)$_c$ and (0 0 14)$_c$ reflections, which are allowed and forbidden, respectively, in the high-temperature cubic phase I (*Fd*–*3m*) (subscript c denotes cubic crystal system; subscripts t and o denote tetragonal and orthorhombic crystal systems, respectively, in the following). To improve the resolution of the diffraction angle $\theta$, X-rays with wavelengths (energies) of 1.25 Å (9.90 keV) and 1.42 Å (8.73 keV) were employed for the (0 0 16)$_c$ and (0 0 14)$_c$ reflections, respectively, so that both reflections appear at high angles with $2\theta > 150°$. In addition, the energy and angular divergences of the incident X-rays were reduced by narrowing the slit before the beamline monochromator. Consequently, we were able to identify a distinct peak split caused by minute tetragonal distortion.

Single crystal samples were obtained by recrystallization of single crystals prepared by chemical vapor transport using the same method described previously.[24] An approximately 2 × 2 × 2 mm$^3$ CRO single crystal was attached to a copper plate and cooled in a $^4$He closed-cycle refrigerator equipped with a four-circle diffractometer. The measurements were performed between 283 and 10 K. To investigate the temperature dependences of the two diffraction angle ranges covering (0 0 16)$_c$ and (0 0 14)$_c$ reflections, separate temperature scans were carried out. The temperature of the sample was monitored by a silicon diode thermometer attached on the copper plate. The error of the sample temperature was estimated to be less than 0.05 K. The scattered X-rays were detected by a two-dimensional pixel array detector (XPAD S70, imXpad, La Ciotat). Because the tetragonal domains of the low-temperature phases are less than a few tens of μm,[17,24] this experimental setup with an exposed area of 1 × 1 mm$^2$ includes reflections from numerous domains. Therefore, the diffraction pattern is a pseudo-powder pattern.

For the intensity fits for diffraction at each temperature, we accounted for the asymmetries of the peaks by introducing skewness into the skew normal distribution function. The asymmetry was already present in the cubic phase, and the peak was slightly elongated to a lower angle than would be predicted by a normal distribution (Gaussian) under ideal conditions. This could be due to the fact that the single-crystal sample was large and diffraction occurred from a certain-width space rather than a point. By fitting the diffraction intensities near the (0 0 16)$_c$ reflection, the location of the peak's centre and the lattice constants were determined. The error of the lattice constants in the fitting is extremely small at 10$^{-6}$ Å, but the actual error must be larger. In addition, the extinction information for each phase was obtained by estimating the diffraction intensities close to the (0 0 14)$_c$ reflection. The phase transition temperature was determined based on the temperature-dependent change in the intensity of the reflection and the lattice distortion.

## 3. Results

### 3.1 Observed four phases with distinct crystal structures

Figure 2 depicts representative pseudo-powder XRD patterns at four temperatures. These diffraction patterns can be distinguished by the difference in number of peaks. At $T$ = 252 K, which corresponds to phase I, the (0 0 16)$_c$ reflection is observed as a single peak, while the (0 0 14)$_c$ reflection is absent, which is consistent with the space group *Fd*–*3m*; the (0 0 14)$_c$ reflection is forbidden by the reflection condition of *h* 0 0: *h* = 4*n* for the *d*-glide symmetry operation. At $T$ = 149 K, which corresponds to phase II, the (0 0 16)$_c$ reflection pattern splits into two peaks, a low-angle peak with low intensity and a high-angle peak with high intensity. Two similar peaks are observed at the location of the $(0\,0\,14)_c$ reflection. In contrast to the previous XRD pattern, in which the diffraction peaks were broad and the two peaks were indistinguishable,[6] this pattern demonstrates the impact of enhanced crystallinity and angle resolution. The appearance of reflections corresponding to $(0\,0\,14)_c$ reflections means that the $d$ glide plane is lost and spatial inversion symmetry is broken. These results are consistent with the tetragonal space group $I\text{–}4m2$.

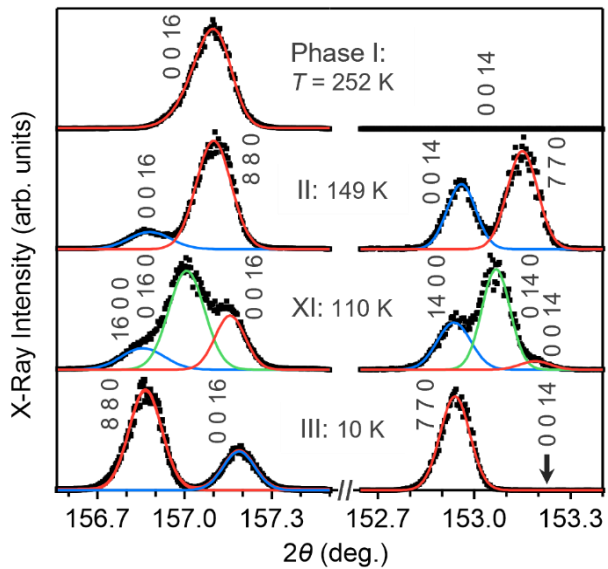

Figure. 2. Typical pseudo-powder XRD patterns from multidomain containing the $(0\,0\,16)_c$ and $(0\,0\,14)_c$ reflections at 252 K (corresponding to phase I), 149 K (phase II), 110 K (phase XI), and 10 K (phase III). The curves that best fit a single Gaussian or multiple skew-normal distribution functions are represented by solid lines. The peak indices are based on cubic, tetragonal, orthorhombic, and tetragonal unit cells, respectively.

The transformation from cubic to tetragonal ($a \neq c$) generally results in the formation of three types of domains, with the $c$-axis of the tetragonal structure facing the three directions [1 0 0], [0 1 0], and [0 0 1] of the cubic structure. In the powder pattern, for instance, the reflections corresponding to $(16\,0\,0)_c$ and $(0\,16\,0)_c$ appear in the same location, while the reflection corresponding to $(0\,0\,16)_c$ appears in a different location. Assuming that the structural change during the phase transition is minimal and the domain orientation is completely random, the ratio of the intensities of the two peaks should be 2:1. In addition, the unit cell vector of a tetragonal structure is expressed as follows during the transition from a face-centered-cubic to a body-centered-tetragonal lattice: $\mathbf{a}_t = (\mathbf{a}_c - \mathbf{b}_c)/2$, $\mathbf{b}_t = (\mathbf{a}_c + \mathbf{b}_c)/2$, $\mathbf{c}_t = \mathbf{c}_c$; the length $a_t$ is $a_c/\sqrt{2}$.

We can assign the following indices to the two peaks of phase II based on this information: $(0\,0\,16)_t$ [$(0\,0\,14)_t$] reflection for the weak low-angle peak and $(8\,8\,0)_t$ [$(7\,7\,0)_t$] reflection for the strong high-angle peak. Noticed, however, that the ratio of intensity between $(0\,0\,16)_t$ [$(0\,0\,14)_t$] and $(8\,8\,0)_t$ [$(7\,7\,0)_t$] reflections deviate from the expected value of 1:2. This is due to the fact that the domain distribution was not completely random. In addition, note that the ratio of intensities varies between the two reflections. This may be due to the fact that the domain distribution was influenced by the temperature history of the cooling process; the two measurements were performed in separate temperature runs. Consequently, tetragonal distortion occurs in phase II when $c_t$ exceeds $\sqrt{2}a_t$. Consistent with previous estimations,[6] the tetragonal strain $2(c_t - \sqrt{2}a_t)/(c_t + \sqrt{2}a_t)$ is 0.050%.

In the $T = 10$ K XRD pattern for phase III, two peaks are observed at the $(0\,0\,16)_c$ reflection location and one at the $(0\,0\,14)_c$ reflection location. The structure is tetragonal due to the splitting of the $(0\,0\,16)_c$ reflection; however, compared to phase II, the intensity of the high and low angle peaks has been reversed, with $(8\,8\,0)_t$ at low angle and $(0\,0\,16)_t$ at high angle. Thus, $c_t$ is less than $\sqrt{2}a_t$, indicating that the tetragonal strain has been reversed at $T_{s2}$. Comparable to phase II, the tetragonal strain in phase III is as low as –0.057%.

The observed peak at the $(0\,0\,14)_c$ reflection location in phase III is determined to be the $(7\,7\,0)_t$ reflection using the lattice constant obtained from the $(8\,8\,0)_t$ reflection. The $(0\,0\,14)_t$ reflection that ought to be observed at 153.23° on the high-angle side is not observed. If we assume the presence of a peak at that location and fit the pattern as a double peak, the $(0\,0\,14)_t$ reflection intensity is merely 0.5% of the $(7\,7\,0)_t$ reflection; therefore, we conclude that there is no $(0\,0\,14)_t$ reflection within the experimental error. This $(0\,0\,14)_t$ reflection is forbidden in $I4_122$ due to the diffraction condition of $0\,0\,l$: $l = 4n$, which is derived from the $4_1$ helical axis. Contrarily, there is no such extinction rule for the $F222$ structure, which was found to exist below 80 K via Raman scattering experiments,[20] thereby ruling out the $F222$ structure. Also disqualified for the same reason is the $I\text{–}4$ structure.

A diffraction pattern clearly distinguishable from phases II and III was observed at $T = 110$ K. Three peaks were observed at both reflection locations, which can be indexed as $(16\,0\,0)_o$ [$(14\,0\,0)_o$], $(0\,16\,0)_o$ [$(0\,14\,0)_o$], and $(0\,0\,16)_o$ [$(0\,0\,14)_o$] from the lowest angle, assuming an orthorhombic unit cell. This new phase is designated as phase XI; to date, ten phases, including the high-pressure phases, have been reported.[3] When the monoclinic angle is close to 90 degrees, it is difficult to determine whether the phase is orthorhombic or monoclinic from this experiment. However, magnetic torque measurements also support the orthorhombic structure.[23] In addition, according to a recent first-principles phonon calculation, the optimal space group for low temperatures is $F222$.[20] As discussed previously in terms of symmetry,[12,17,18] phase XI, which is continuously connected to $I$–$4m2$ and $I4_122$, is likely $F222$. The characteristics of each phase are summarized in Table I.

## 3.2 Structural change across $T_{s1}$

The temperature variation of the XRD profile near $T_{s1}$ is shown in Fig. 3(a). At high temperatures, the (0 0 16)$_c$ reflection is a single peak and shifts to the high-angle side upon cooling due to lattice contraction. At approximately 200 K, however, it shifts to the low-angle side, indicating an atypical lattice expansion. At the same time, a low-angle side broadening is observed, and at 149 K, the peak completely separates into (0 0 16)$_t$ and (8 8 0)$_t$ reflections, as shown in Fig. 2.

Figure 3(b) illustrates the temperature dependence of the half width at half maximum (FWHM) obtained by fitting a single Gaussian function to the data. The FWHM remains constant above 202 K and rises close to 201 K. The evolution of lattice constants was determined by fitting double skew-Gaussian functions to the diffraction patterns below 201 K. The deviation from the cubic crystal $\sqrt{2}a_t - c_t$, which is regarded as an OP, exhibits a temperature dependence consistent with a second-order transition [Fig. 3(b)]. By fitting the data between 201 K and 188 K to a power function $(T_c - T)^\beta$, a critical temperature $T_c = T_{s1} = 201.5(1)$ K and a critical exponent $\beta = 0.34(1)$ were determined. When combined with the increase in FWHM at approximately the same temperature, $T_{s1}$ can be considered the transition temperature to the tetragonal structure.

As shown in Fig. 3(a), the peak corresponding to the forbidden (0 0 14)$_c$ reflection appears at 201 K and grows in intensity and width as temperature decreases. The peak eventually separates into the (0 0 14)$_t$ and (7 7 0)$_t$ reflections at 149 K (Fig. 2). To obtain the temperature dependence of the intensities of the two reflections, the patterns are fitted with two skewed Gaussians. As shown in Fig. 3(b), the (0 0 14)$_t$ reflection increases with $T_c = 201.23(7)$ and the critical exponent $\beta = 0.42(2)$ in the form $(T_c - T)^\beta$, while the (7 7 0)$_t$ reflection exhibits a different temperature dependence with the same $T_c$; this difference in temperature dependence may be due to anisotropic atomic shifts. Recent XRD study that focused on superlattice reflections uncovered anisotropic atomic displacements.[14] $T_c$ is nearly identical to $T_{s1}$ according to the $\sqrt{2}a_t - c_t$ plot of Fig. 3(b). We therefore conclude that the temperatures at which tetragonal distortion and $d$-glide loss occur are equivalent within the experimental error margin. Although it is possible that a cubic $F$–$43m$ phase exists between phases I and II in which only spatial inversion is broken, as predicted by the primary OP of $A_{2u}$ rather than $E_u$,[12,17,18] our results exclude this possibility.

Table 1. Structural parameters for the series of ambient-pressure phases of Cd$_2$Re$_2$O$_7$. Lattice constants are determined from the location of the peak's centre near the (0 0 16)$_c$ reflection.

| Phase | I | II | XI | III |
|---|---|---|---|---|
| Transition temp. | – | $T_{s1} = 201.5(1)$ K | $T_{s2} = 115.4(1)$ K | $T_{s3} \sim 100$ K |
| Space group | $Fd$–$3m$ | $I$–$4m2$ | $F222$ | $I4_122$ |
| Observed reflections | 0 0 16 | 7 7 0, 0 0 14 <br> 8 8 0, 0 0 16 | 14 0 0, 0 14 0, 0 0 14 <br> 16 0 0, 0 16 0, 0 0 16 | 7 7 0 <br> 8 8 0, 0 0 16 |
| Lattice constants | $a_c = 10.2209$ Å <br> (210 K) | $a_t = 7.2283$ Å <br> $c_t = 10.2274$ Å <br> (130 K) | $a_o = 10.2274(1)$ Å <br> $b_o = 10.2243(2)$ Å <br> $c_o = 10.2215(3)$ Å <br> (110 K) | $a_t = 7.2314$ Å <br> $c_t = 10.2209$ Å <br> (10 K) |
| Tetragonal distortion: <br> $2(c_t - \sqrt{2}a_t)/(c_t + \sqrt{2}a_t)$ | – | 0.050% <br> (130 K) | – | −0.057% <br> (10 K) |
| $E_u$ order parameter | (0, 0) | (0, $\eta_2$) | ($\eta_1$, $\eta_2$) | ($\eta_1$, 0) |

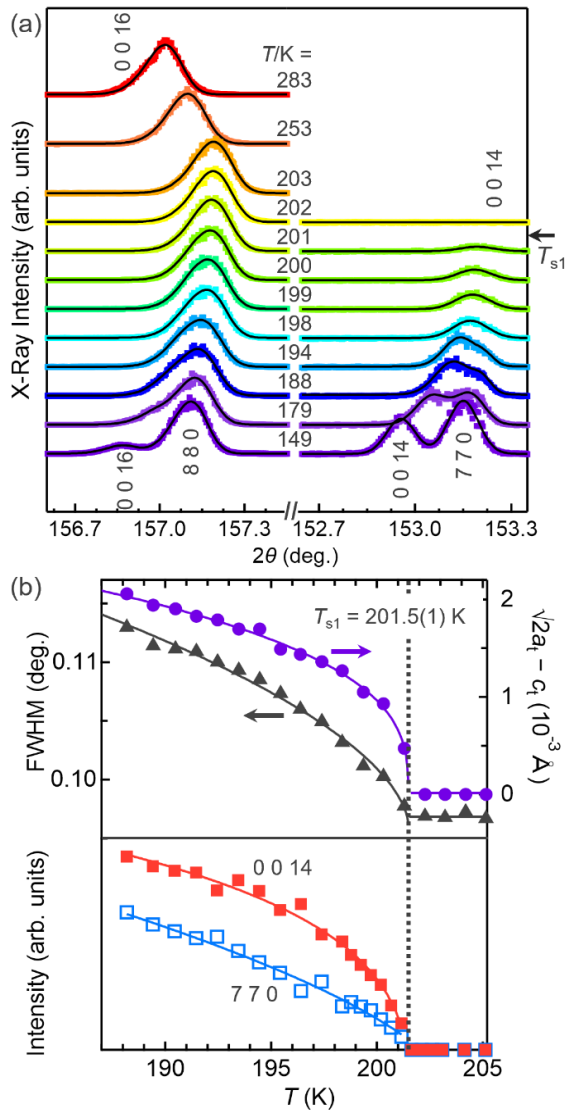

Figure 3. (a) Temperature evolution of the XRD patterns across $T_{s1}$ near the $(0\,0\,16)_c$ and $(0\,0\,14)_c$ reflections of phase I. The solid lines represent single Gaussian function fits above 202 K and double skew Gaussian function fits below 201 K. (b) Temperature dependences close to $T_{s1}$ of the FWHM and the tetragonal distortion $\sqrt{2}a_t - c_t$ of the $(0\,0\,16)_c$ (top), as well as the intensities of the $(7\,7\,0)_t$ and $(0\,0\,14)_t$ reflections (bottom). Solid lines are fitted to the form $(T_c - T)^\beta$. The error bars of the standard deviations of the skew-Gaussian fits are too small to observe.

### 3.3 Structural changes around $T_{s2}$ and $T_{s3}$

The changes close to $T_{s2}$ are depicted in Fig. 4(a). Above 117 K, which corresponds to phase II, there is no change in the shape of the two peaks in the $(8\,8\,0)_t/(0\,0\,16)_t$ reflection. Below this temperature, however, only the high-angle $(8\,8\,0)_t$ reflection broadens as the temperature decreases, and at 111 K, it clearly separates into two peaks. The high-angle peak remains at almost the same position, whereas the low-angle peak shifts to a lower angle. At approximately 100 K, the $(0\,0\,16)_t$ reflection gradually shifts to the high-angle side and merges with the central peak, resulting in two peaks once more; this temperature is denoted as $T_{s3}$. Consequently, a distinct phase XI with an apparent orthorhombic structure exists between $T_{s2}$ and $T_{s3}$ (Fig. 2).

If this change were the result of a single first-order transition, phases II and III would coexist within this temperature range. The pattern should then be a simple addition of the 89 K and 117 K patterns, with only their relative intensities varying with temperature. However, the continuous peak shift observed in Fig. 4(a) disproves the coexistence of the two phases. Similar continuous changes have also been observed in magnetic torque measurements[23] and Cd NMR spectra.[22] Therefore, the change in this region is not the result of a first-order transition, but rather two successive second-order transitions.

The $(7\,7\,0)_t/(0\,0\,14)_t$ reflection exhibits essentially the same temperature dependence as the $(8\,8\,0)_t/(0\,0\,16)_t$ reflection in the vicinity of $T_{s2}$, but below $T_{s3}$ it transforms into a single peak corresponding to the $(7\,7\,0)_t$ of phase III. The intensity of the $(0\,0\,14)_o$ reflection in phase XI decreases as the temperature decreases from $T_{s2}$ and disappears below 100 K. The $(0\,0\,14)_o$ reflection intensity in Fig. 4(b) decreases rapidly below $T_{s2}$, then approaches zero asymptotically near $T_{s3}$. However, determining $T_{s3}$ from this temperature dependence of the $(0\,0\,14)_o$ reflection intensity is difficult.

The temperature dependence of the lattice constants was determined by fitting triple skew-Gaussian functions to the XRD pattern between 120 K and 90 K [Fig. 5(a)]. In phase XI, $\sqrt{2}a_t$ from phase II splits into two ($b_o$, $c_o$), while $c_t$ from phase II becomes $a_o$. In phase III, $a_o$ and $b_o$ approach and transform into $\sqrt{2}a_t$.

Figure 4(b) depicts the temperature dependences of two orthorhombic distortion types, $d_1 = b_o - c_o$ and $d_2 = a_o - b_o$. $d_1$ develops rapidly from $T_{s2}$ and exhibits an OP-like behavior for a second-order transition. The fitting of the power function yields a transition temperature of $T_{s2} = 115.4(1)$ K and a critical exponent of $\beta = 0.51(3)$. Unlike $d_1$, $d_2$ decreases as $T_{s3}$ approaches and approaches zero asymptotically. A similar power function fit yields $T_{s3} = 99.9(7)$ K and $\beta = 1.5(1)$. This transition temperature is identical to the temperature at which the $(0\,0\,14)_o$ reflection vanishes.

This peculiar temperature dependence raises the question of whether $T_{s3}$ is a crossover rather than a phase transition. If this is the case, the coincidence of $a_o$ and $b_o$ is coincidental, and phase III does not exist at the lowest temperature, but phase XI does. Nevertheless, the observed disappearance of the $(0\ 0\ 14)_o$ reflection below $T_{s3}$ indicates a change in space group or symmetry, proving the existence of a phase transition at $T_{s3}$. We conclude that phase XI is a new orthorhombic phase that exists within a 15 K temperature window between 115 K and 100 K.

$a_o - b_o$ (red triangles) as well as the $(0\ 0\ 14)_t$ reflection intensity. The data for $d_1$ and $d_2$ are fitted by solid lines to $(T_c - T)^\beta$: $(T_c, \beta)$ = [115.4(1), 0.51(3)] and [99.9(7), 1.5(1)], respectively.

### 3.4 Lattice constants and cell volume

Figure 5 illustrates the temperature dependence of the lattice constants and volume over the entire temperature range, as determined by diffraction intensity fits near the $(0\ 0\ 16)_c$ reflection location. The temperature dependence of the lattice constants near $T_{s1}$ replicates previous observations.[6,7] At low temperatures, there are significant changes near $T_{s2}$ and $T_{s3}$ with a reversal of tetragonal distortion between $c_t > \sqrt{2}a_t$ above $T_{s2}$ and $c_t < \sqrt{2}a_t$ below $T_{s3}$, in contrast to previous reports in which the change was always smooth with $c_t > \sqrt{2}a_t$ down to 10 K.[6,7]

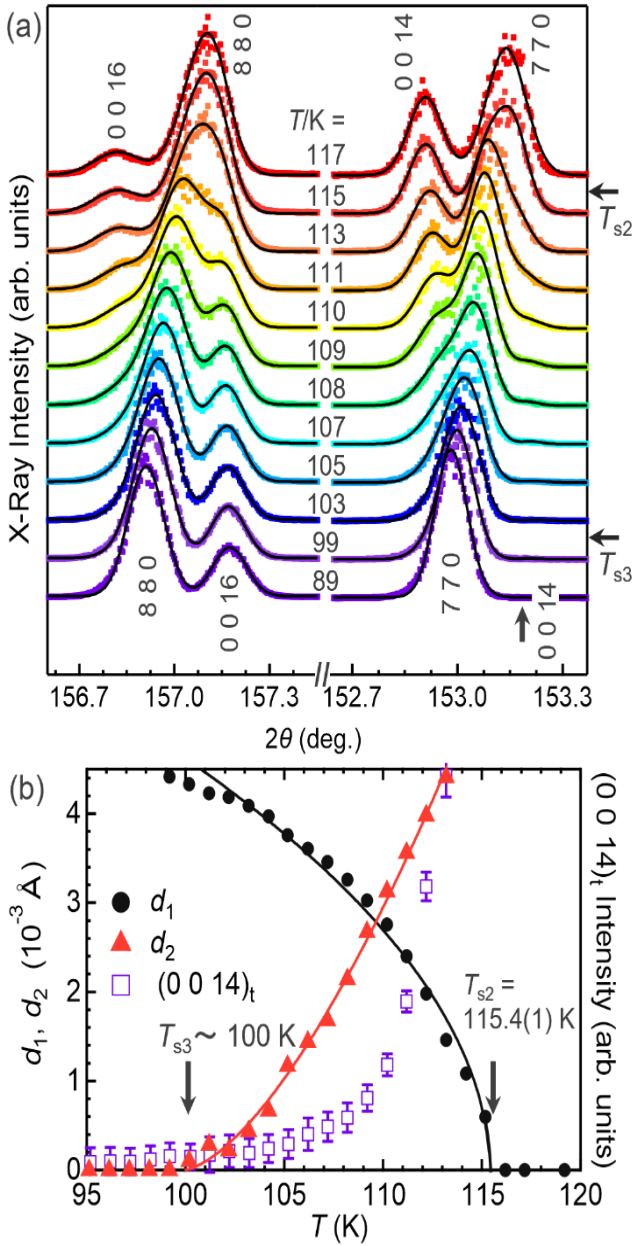

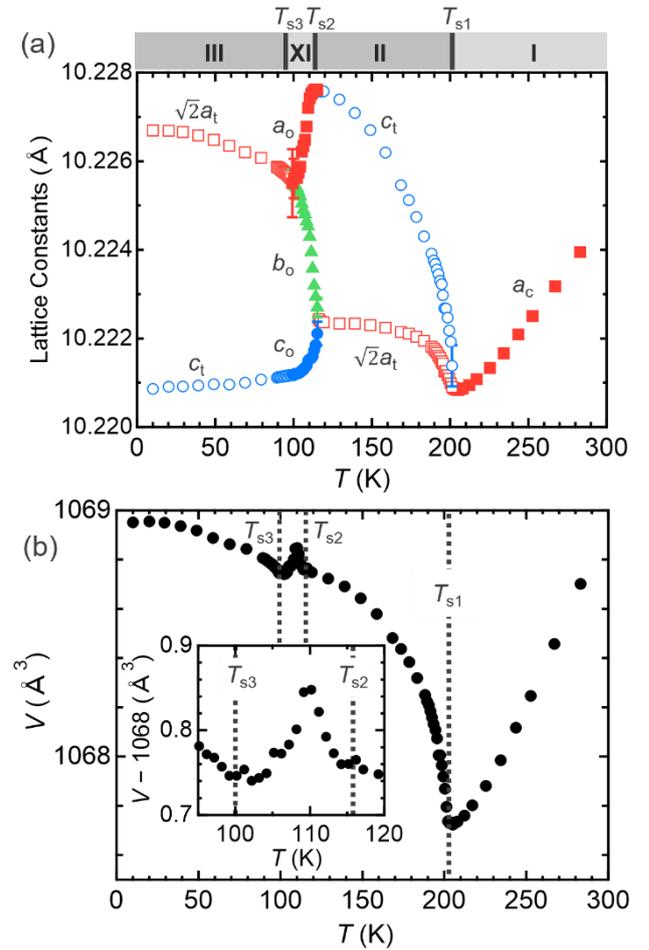

Figure 4. (a) Temperature evolution of the $(8\ 8\ 0)_t/(0\ 0\ 16)_t$ and $(7\ 7\ 0)_t/(0\ 0\ 14)_t$ reflections in the temperature window encompassing $T_{s2}$ and $T_{s3}$. Solid lines represent triple skew-Gaussian function fits. (b) Temperature dependences of the orthorhombic distortions $d_1 = b_o - c_o$ (black circles) and $d_2 =$

Figure 5. Temperature dependences of (a) lattice constants and (b) unit cell volume for CRO. The error bar for the lattice constants is negligible except near the transition temperatures. The inset of (b) enlarges the volume change at phase XI.

In phase I, as shown by the temperature dependence of unit cell volume in Fig. 5(b), the lattice thermally contracts upon cooling like normal materials, whereas in phases II and III, the volume expands and becomes even greater than that at room temperature. In contrast, the volumes of the pyrochlore oxides $Tl_2Mn_2O_7$ and $Lu_2V_2O_7$ decrease monotonically with decreasing temperature, by 0.64% (14 K) and 0.42% (59 K) from room temperature, respectively.[26] The negative thermal expansion of CRO at low temperatures may be the hallmark of phase transitions driven by the energy stabilization of the electronic system. The volume reaches a peak in phase XI, but the reason for this is unknown. It is possible that this is an artifact of the analysis, as it was difficult to distinguish between the three peaks. Alternatively, it could reveal an intriguing characteristic of phase XI.

## 4. Discussion

### 4.1 Possible space group for phase XI

The OP of the I–II–III successive phase transitions is understood in terms of a two-dimensional irreducible representation of $E_u$ at the Γ point of the Brillouin zone.[12] The $E_u$ OP is represented by the two-dimensional vector $\boldsymbol{\eta} = (\eta_1, \eta_2)$. In phase II of space group $I\bar{4}m2$, only $\eta_2$ has a finite value, whereas in phase III of space group $I4_122$, only $\eta_1$ has a finite value (Table 1).

In addition to $I\bar{4}m2$ and $I4_122$, $F222$ is a candidate for a structure with low symmetry that can be transferred by the $E_u$ OP from the space group $Fd\bar{3}m$.[12,17,18] The OP of $F222$ is their combination; $\boldsymbol{\eta} = (\eta_1, \eta_2)$. It is therefore reasonable to assume that phase XI, which is continuously linked to phases II and III, possesses an $F222$ structure. The three peaks seen in phase XI provide support for this orthorhombic structure.

### 4.2 Polar coordinate description of the OPs

Assuming $\eta_1 = \eta\cos\theta$ and $\eta_2 = \eta\sin\theta$, the OP can be represented in polar coordinates as $\boldsymbol{\eta} = (\eta, \theta)$, as shown in Fig. 6(a);[27] $\eta$ and $\theta$ represent the amplitude and phase of the OP, respectively. Since the vector $\boldsymbol{\eta}$ has sixfold rotational symmetry in the $O_h$ point group, there are six equivalent states for each rotation of 60 degrees. These correspond to six distinct domains with various $c$-axis orientations in an induced tetragonal structure. In this polar coordinate plane, the OP vectors of $I\bar{4}m2$ and $I4_122$ are located on the lines $\theta = k\pi/6$ with $k = 2n + 1$ and $2n$, respectively, and the area between these two lines represents $F222$.

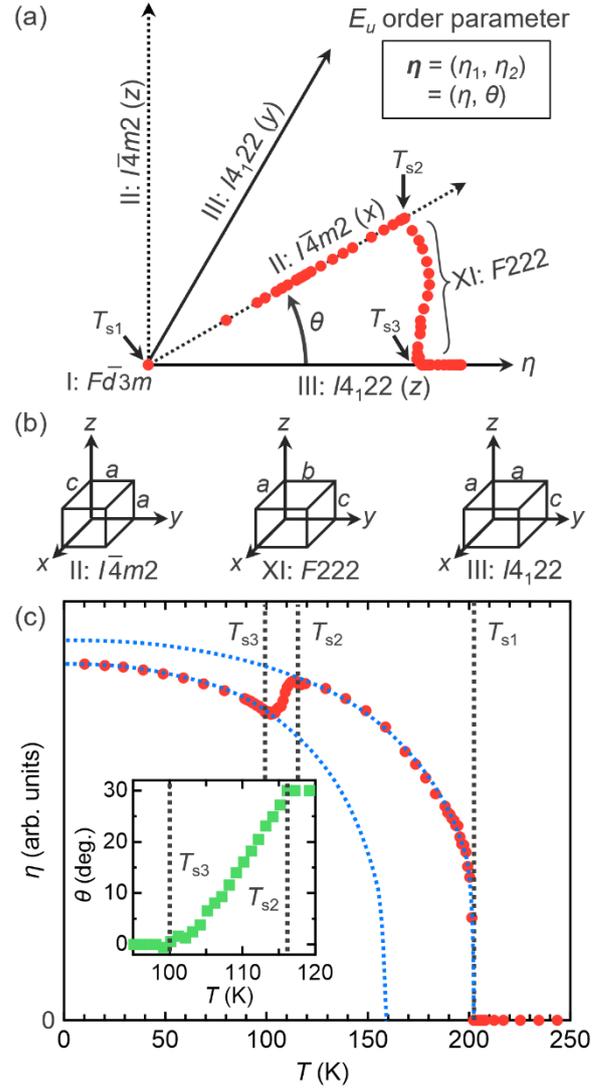

Figure 6. (a) Polar coordinate representation of the $E_u$ OP $\boldsymbol{\eta} = (\eta, \theta)$. Phases I, II, III, and XI are located, respectively, at the origin, on the $\theta = n\pi/3$ lines (dashed lines), on the $\theta = (2n + 1)\pi/6$ lines (solid lines), and between them. In the explanation of the text, the $\theta = 30°$ and $0°$ lines are selected for phase II with its $c$ axis along $x$, and for phase III with its $c$ axis along $z$, respectively. The red marks represent the OPs that are temperature-dependent. (b) Relationship between the unit cell axes of the three phases studied in the text. (c) Temperature dependences of the amplitude $\eta$ (main panel) and phase $\theta$ (inset). The blue dashed lines serve as eye guides.

Consider the variation in lattice constants caused by $E_u$ distortion with respect to the coordinate axes ($x$, $y$, $z$) of an orthorhombic structure. Taking into account $\eta$ up to the second order in the free energy, the lattice constants of the orthorhombic unit cell $a_o$, $b_o$, and $c_o$ in each direction are as follows:[27]

$$a_o = a_0 + a_A \eta^2 + a_E \eta^2 \cos(2\theta + 2\pi/3),$$
$$b_o = a_0 + a_A \eta^2 + a_E \eta^2 \cos(2\theta - 2\pi/3), \quad (1)$$
$$c_o = a_0 + a_A \eta^2 + a_E \eta^2 \cos(2\theta),$$

where $a_0$ is a temperature-dependent constant due to thermal expansion, and $a_A$ and $a_E$ are temperature-independent constants.

In $I\text{–}4m2$, for example, substituting $\theta = \pi/6$ results in the expression $b_o = c_o$, which represents the $x$ domain as $c_t // x$. The expected domain of $I4_122$ transforming from this domain is the one that minimizes the change in OP, therefore $\theta = 0$ or $\theta = \pi/3$. $\theta = 0$ yields $a_o = b_o$, which is the $z$ domain with $c_t // z$, while $\theta = \pi/3$ yields the $y$ domain. Figure 6(b) depicts the relationship between the axis orientations of $\theta = 0$ ($I\text{–}4m2$), $\pi/6$ ($I4_122$), and an intermediate orientation for $F222$. This domain transition with a 90° rotation of the $c_t$ axis during the transition between phases II and III was actually observed by polarized light microscopy.[24] This is thought to be associated with tetragonal distortion switching.

Transforming equation (1), we obtain

$$\tan(2\theta) = \sqrt{3}(b_o - a_o)/(2c_o - a_o - b_o), \quad (2)$$
$$\eta = [(2c_o - a_o - b_o)/3a_E \cos(2\theta)]^{1/2}.$$

We determined the temperature dependences of magnitude $\eta$ and phase $\theta$ of the OP by substituting the experimentally obtained lattice constants into equation (2) [Fig. 6(c)]. $\eta$ develops rapidly at $T_{s1}$, indicating that it is indeed the primary OP of the phase transition. Fixing the transition temperature to 201.5 K as mentioned above and fitting the data with a power function, the critical exponent $\beta = 0.223(5)$ reproduces the data close to $T_{s1}$. In contrast, $\eta$ decreases once at $T_{s2}$ and increases again below $T_{s3}$. This suggests that the hypothetical transition temperature of phase III in the absence of phases II and XI could be lower than that of phase II, resulting in a smaller OP development in phase III at these temperatures. Nevertheless, since the observed decrease in $\eta$ below $T_{s2}$ is uncommon for a phase transition with a single OP, a secondary OP may play a significant role in these complex phase transitions. On the other hand, $\theta$ varies almost linearly between 30° and 0° in phase XI [Fig. 6(c), inset].

Figure 6(a) displays the variation in polar coordinates of OP as a function of temperature. Phase I is located at the origin with zero amplitude. When the transition from phase I to phase II occurs at $T_{s1}$, $\eta$ becomes finite and increases along the line $\theta = 30°$ with decreasing temperature. At $T_{s2}$, during the transition to phase XI, $\theta$ begins to approach zero while $\eta$ slightly decreases. When phase III is reached at $T_{s3}$, the OP develops along the $\theta = 0°$ line. A two-dimensional $E_u$ OP can therefore explain the continuous phase transition of CRO, including phase XI.

The linear variation in $\theta$ between $T_{s2}$ and $T_{s3}$ suggests that $\theta$ is the essential parameter for describing phase transitions in this region. This contrasts with typical phase transitions, such as the $T_{s1}$ transition, in which the magnitude of OP increases as the temperature decreases. According to Landau theory, this change in $\theta$ is caused by the dominance of a term of extremely high order (12th order).[27] The higher order term dominates the $T_{s2}$–$T_{s3}$ transition, probably because phases II and III are nearly degenerate in energy. However, from a symmetry perspective, the two cannot be connected continuously; there should be a jump between them. To avoid this, it is believed that the $F222$ phase described by the linear coupling of both OPs will intervene as phase XI.

### 4.3 Remarks to the previous experiments at around $T_{s2}$

The transition at ~120 K has been considered to be a first-order transition.[3] This is primary due to the small temperature hysteresis observed in the electrical resistivity (at 120–114 K for crystal 1A and at 117–112 K for crystal 40A in reference 3) and the sharp peak observed at 112 K in the heat capacity measurement.[3] However, there was significant sample dependence in these data. The electrical resistivity of particular crystals exhibited no hysteresis. In fact, it was not observed in single crystals of comparable quality to that used in this study. In addition, heat capacity measurements revealed that the second crystal in reference 3 exhibited not a sharp peak, but rather two broad bumps near 112 K and 100 K;[3] these temperatures are close to $T_{s2}$ and $T_{s3}$.

Phase XI is believed to be the result of a delicate balance between phases II and III. In addition, $T_{s3}$ is a transition of higher order, as indicated by its peculiar temperature dependence, and thermal equilibrium may take a considerable amount of time to achieve; as stated previously, $T_{s3}$ is not a crossover because the symmetry between phases XI and III is distinct. In addition, the elastic energy associated with domain formation and domain wall pinning as a result of defects may be in conflict with thermodynamic equilibrium. Consequently, the behavior near $T_{s2}$ and $T_{s3}$ may be highly dependent on sample quality and measurement conditions and techniques. Accordingly, we believe that the two successive transitions observed in this study are consistent with prior experimental findings and are an intrinsic property of CRO.

Recent magnetic torque measurements on high-quality crystals confirmed the existence of a second-order phase transition at 115 K and suggested the existence of a transition at a lower temperature.[23] Takigawa's Cd-NMR experiments revealed the existence of a phase with lower symmetry than the tetragonal one between 115 and 100 K,[22] which roughly corresponds to $T_{s2}$ and $T_{s3}$. In light of our structural data, we conclude that the three-step sequential phase transitions in $E_u$ OP have been experimentally evidenced.

*4.4 Characteristics of the multipolar transitions of CRO*

A possible origin of the phase transitions in CRO is the Fermi liquid instability of SOCM. Spin-degenerate Fermi surfaces with large SOCs are stabilized by spontaneous ISB, forming a multipole order with spin-split Fermi surfaces. Based on crystal symmetry considerations, phases II and III correspond to $x^2 - y^2$- and $3z^2 - r^2$-type ETQ orders, respectively.[13] Phase XI, whose OP is represented by a linear combination of the OPs of phases II and III, is another entangled odd-parity multipole.

To experimentally establish ETQ orders in CRO, it is required to confirm the presence of the secondary OP of even-parity $E_g$. According to group theoretical considerations, the electric quadrupole (EQ) order corresponding to $E_g$ should coexist with the ETQ order resulting from $E_u$.[13,17,18] Unfortunately, the accuracy of our structural data was insufficient to detect the coexistence of OPs other than $E_u$. Nevertheless, magnetic torque measurements, which are sensitive to even parity OPs, revealed that odd-parity $E_u$ and even-parity $E_g$ actually coexist.[23] On the other hand, SHG measurements, which are sensitive to odd parity OPs, revealed that $T_{2u}$, $T_{1g}$, and $E_u$ coexist, with $E_u$ being a secondary OP.[17] However, it has been proposed that $E_u$ alone can reproduce the same angular dependence of the signal observed in SHG measurements for the $T_{2u}$ mode.[17,18] Consequently, the previous experimental findings can be explained by a scenario assuming primarly $E_u$ and secondary $E_g$. In the future, the existence of OPs other than $E_u$ may be clarified using ultrasonic measurements and other suitable measurement techniques.

The emergence of as many as three electronic multipole phases in CRO may be a manifestation of the diversity of Fermi liquid instability resolution mechanisms in SOCM. In structural phase transitions caused by electronic instabilities, such as the Jahn–Teller effect, two types of states can frequently appear.[28] For instance, stabilizations of the *d*-orbital energy by stretching and contracting the octahedron are degenerate. However, once a structural transition occurs at low temperatures and one deformation is chosen, the structure stabilizes with a large structural deformation, and subsequent temperature changes rarely cause the other deformation to appear. This is due to large electron–phonon interactions. In contrast, the exceptional sequential occurrence of multiple structural deformations in CRO is most likely due to the weak electron–phonon interactions, and the phase transition is solely governed by the competition of electron system energies. Another aspect of SOCM's Fermi liquid instability may be the presence of up to seven phases under high pressure.[9] CRO is the ideal compound for studying pure electronic phase transitions in SOCM.

**4. Conclusion**

High-resolution synchrotron radiation XRD experiments were performed on a high-quality single crystal of a SOCM candidate CRO to clarify the details of the structural changes, including the temperature dependence of the lattice constants. The transition around 120 K, previously believed to be a first-order transition, consists of two successive continuous transitions with phase XI of the space group $F222$ in between. The three successive phase transitions of CRO can be understood in a unified manner in terms of the two-dimensional order parameter $E_u$: $T_{s1}$ is a typical phase transition where the amplitude of the OP develops, whereas $T_{s2}$ and $T_{s3}$ are exceptional phase transitions where the phase of the OP changes. Negative thermal expansion below $T_{s1}$ and the transition of increased symmetry below $T_{s3}$ indicate that the transitions are exclusively electronic in nature. Phases II and III are ETQ odd-parity multipole orders of the $x^2 - y^2$ and $3z^2 - r^2$ types, respectively, and phase XI is regarded as their superposition state. The appearance of these multipole phases must reflect the instability of Fermi liquids in SOCMs.

**Acknowledgements**

The authors are grateful to J. Yamaura, M. Takigawa, S. Uji, Y. Yokoyama, and M. Mizumaki for insightful discussion. They appreciate Y. Motome and H. Kusunose for their helpful comments. They also thank H. T. Hirose for the ETQ images in Fig. 1. This work was performed under the approval of the Photon Factory Program Advisory Committee (Proposal No. 2020G628). This research was supported by Japan Society for the Promotion of Science (JSPS) KAKENHI grant numbers 20H01858, 22H04462, and 22H01178.